# Ancient Greek Technology: An Immersive Learning Use Case Described Using a Co-Intelligent Custom ChatGPT Assistant

Vlasios Kasapakis, and Leonel Morgado, *Member, IEEE Education Society*

*Abstract*— Achieving consistency in immersive learning case descriptions is essential but challenging due to variations in research focus, methodology, and researchers' background. We addresses these challenges by leveraging the Immersive Learning Case Sheet (ILCS), a methodological instrument to standardize case descriptions, that we applied to an immersive learning case on ancient Greek technology in VRChat. Research team members had differing levels of familiarity with the ILCS and the case content, so we developed a custom ChatGPT assistant to facilitate consistent terminology and process alignment across the team. This paper constitutes an example of how structured case reports can be a novel contribution to immersive learning literature. Our findings demonstrate how the ILCS supports structured reflection and interpretation of the case. Further we report that the use of a ChatGPT assistant significantly supports the coherence and quality of the team members development of the final ILCS. This exposes the potential of employing AI-driven tools to enhance collaboration and standardization of research practices in qualitative educational research. However, we also discuss the limitations and challenges, including reliance on AI for interpretive tasks and managing varied levels of expertise within the team. This study thus provides insights into the practical application of AI in standardizing immersive learning research processes.

## I. Introduction

Consistency in immersive learning case descriptions is an aspiration of researchers. Achieving this consistency would enable researchers to effectively contrast and compare cases, ultimately leading to broader insights and generalizable findings [1]. By 'contrasting and comparing', we mean examining differences in methodologies, implementations, and outcomes, such as how different learning approaches are deployed or not. Documenting these cases is challenging due to inconsistencies in how researchers present or omit details across studies [2], influenced by different research foci, terminology choices, and diverse researcher backgrounds, often leading to subjective interpretations. To address these challenges, the Immersive Learning Case Sheet (ILCS) is a recently introduced methodological instrument to standardize case descriptions, seeking to enhance comparability and research quality in immersive learning research.

We demonstrate the ILCS effectiveness by presenting an educational case on ancient Greek technology developed in the VRChat virtual world, chosen for its rich educational value and convenience, having been exploited beforehand, within the REVEALING project, on the advantages of using virtual reality (VR) for synchronous learning [3]. This example shows how ILCS brings coherence to case documentation and contributes with the structured case report itself.

Our research team had diverse skill sets—some knew the case content well, while others were familiar with ILCS and its method. We explored the use of a custom ChatGPT assistant to bridge these differences. Approaching it from a co-intelligence perspective [4], where AIs and humans influence mutually in a cognitive ecosystem [5], the custom ChatGPT assistant prompted for more details, provided automated checks on framework compliance, suggested terminology, and supported adherence to ILCS guidelines, facilitating consistency throughout the project.

This paper explores how the ILCS method provides new insights into immersive learning cases and highlights the role of AI-driven tools in enhancing research quality and analysis. By introducing the custom GPT-based assistant and demonstrating its functionality via application to a learning case, this work aids in standardizing immersive learning case descriptions, facilitating future comparative analyses across related studies and projects. AI facilitated consistency checks and real-time suggestions, contributing to more standardized outcomes, but occasional deviations also posed risks to compliance and standardization. We provide insights into the challenges and opportunities encountered in this process.

These arguments are presented as follows: Section 2 covers the background, including immersive synchronous collaborative learning and the ILCS method, and details the research & technology context, Section 3 explains methods, Section 4 provides the case study using the ILCS, followed by the discussion in Section 5, and conclusions in Section 6.

## II. Background

### A. Immersive VR as a medium for synchronous collaborative learning

The integration of immersive VR into education has received increased attention over the last decade due to technological advancements, which have made Head-Mounted Displays (HMDs) accompanied by sophisticated motion controllers more affordable [6]. Initially, immersive VR in education focused on single-user immersive learning environments. However, as technology progressed, social immersive VR platforms enabled remote synchronous collaboration between students and teachers in interactive learning environments, enhancing learning experiences and demonstrating VR's educational potential [7].

Collaborative immersive learning environments have since been developed across disciplines, allowing students to interact, present themselves through avatars, and communicate using body language and voice. In geography education, for instance, VR is used to present 3D representations of real

V. Kasapakis is with the University of the Aegean, Mytilene, Greece (e-mail: v.kasapakis@aegean.gr).

L. Morgado is with Universidade Aberta, Coimbra Delegation, Portugal, and with INESC TEC, Porto, Portugal (e-mail: Leonel.Morgado@uab.pt).

locations, allowing different approaches such as observing virtual locations or engaging in interactive tasks [8]. The ability of VR to present accurate three-dimensional representations and interactions has also led to its adoption in demanding technical specialties such as medical education, where it provides safe environments for surgical training, including collaboration in complex surgeries [9].

Despite the promising cases and results of VR in education, inconsistencies in instructional approaches and contexts make it difficult to compare learning environments and outcomes. Authors often categorize use cases without justifying their decisions with wider taxonomies or frameworks, and descriptions of implementation, pedagogy, and design vary significantly, posing challenges for adopting guidelines and use, by instructors, researchers and industry practitioners [10].

*B. Analysis and Planning of Educational Activities - The Immersive Learning Case Sheet method*

Effective analysis and planning of educational activities in immersive learning environments require a structured method to capture and organize key elements of these activities. However, this process is far from straightforward: there are few instructional design recommendations or frameworks [11], and the research field of immersive learning research is fragmented, limiting the current understanding of how to compare instructional approaches and draw conclusions towards practice [12]. The ILCS Method addresses these challenges by systematically describing and interpreting immersive learning cases, facilitating the planning of new activities and enhancing the analysis of existing ones [13].

The ILCS method combines frameworks: the Immersion Cube conceptual dimensions proposed by Nilsson et al. [14], and the Immersive Learning Brain (ILB) mapping of the literature by Beck et al. [12]. The Immersion Cube focuses on classifying immersion along three conceptual dimensions: system immersion, which refers to feeling present within the environment; narrative immersion, which is linked to the storyline or context; and agency immersion, which relates to the learner's ability to act within the environment (Nilsson et al. referred to the latter as 'challenge' but the ILCS method paper renamed it 'Agency'). The Immersion Cube helps categorize the immersive experience in a conceptual space, and Beck et al. [15] mapped into it the literature-reported uses of immersive learning, offering a broader framework for understanding and comparing learning cases. This mapping is called the Immersive Learning Cube (ILC). Plotting their own immersive learning cases within the ILC, and identifying their practices and strategies in the ILB, educators can locate their cases in relation to other cases in the literature and proximal uses, practices, and strategies, allowing for comparisons and insights into how a case aligns with or diverges from others.

The ILB and ILC frameworks employ a three-level taxonomy to describe educational activities:

- *Uses*: Operational activities without explicit pedagogical rationale or intents lacking specific means for implementation.
- *Practices*: activities with an explicit pedagogical rationale, within specific contexts or limited scope.
- *Strategies*: broader instructional philosophies or plans, that guide learning designs or emerge from them.

The steps for creating the ILCS are divided into two parts: first using the ILB, then applying the ILC. Using the ILB, the steps involve:

1. *Case Description*: Develop a detailed description of the immersive learning activity, focusing on how immersive features are employed pedagogically.
2. *Cluster Selection*: Identify the most relevant ILB clusters based on the educational activity's core attributes.
3. *Tagging of Practices and Strategies*: Compare the activity with the ILB clusters and tag the relevant practices and strategies.
4. *Rewriting*: Refine the case description to acknowledge the tagging.

After completing the ILB part of the Case Sheet process, the ILC is used, involving these steps:

1. *Case Description*: Develop a description that clarifies the relevance of immersion across the conceptual dimensions of system, narrative, and agency, as defined by the Immersion Cube.
2. *Dimensioning and Mapping*: Assess the case's reliance on these dimensions and map it within the ILC, considering how it compares to other known uses.
3. *Locating*: Measure the Euclidean distance between the case and other known uses of immersive learning within the ILC.
4. *Rewriting*: Revise the case description based on its proximity to known uses, clarifying relevant immersion aspects across the conceptual dimensions.

By applying this method, as demonstrated in this paper, educators and researchers can create an ILCS that supports the planning of educational interventions and the comparison of learning outcomes across immersive learning environments. This structured approach seeks to enhance consistency and facilitate meta-analyses of immersive learning cases.

*C. Context: VRChat & the REVEALING project*

The objective of the REVEALING project (https://revealing-project.eu/) was to design, develop, and evaluate collaborative, immersive learning environments that can be integrated into curricular activities organized by Higher Education Institutions (HEIs), presenting teaching materials in an engaging and interactive way while providing meaningful learning experiences. These immersive learning environments are created in close collaboration with HEI teaching staff and VR application developers, utilizing the VRChat platform (https://hello.vrchat.com/) and Meta Quest 2 Head-Mounted Displays (HMDs). VRChat is a platform designed for VR experiences, where users can build and explore immersive worlds, while interacting in real-time with others through personalized avatars. It enables communication using voice, gestures, and spatial interactions. This setup enables students and teachers visiting the immersive learning environments of the REVEALING project to coexist in an immersive virtual space, each controlling their own avatar, communicating via voice chat and body language, and interacting with the virtual environment using the HMD's motion controllers.

### III. METHODS

To facilitate creation of the ILCS, we developed a custom GPT-based assistant, the "Immersive Learning Case Sheet assistant" (ILCS Assistant): https://chatgpt.com/g/g-JDLJLXin5-immersive-learning-case-sheet-assistant.

The ILCS Assistant GPT was provided with several academic preprints as its knowledge base: the Immersive Learning Case Sheet method, along with its associated definitions of educational uses, practices, and strategies [13]; the ILB framework and its list of strategies and practices, their definitions, and the clusters grouping them [12]; the ILC framework, with the description of each Educational Use and its coordinates on the ILC [15]. We also provided the Assistant with a publicly available slideshow summarizing the ILCS method (https://pt.slideshare.net/slideshow/describing-and-interpreting-an-immersive-learning-case-with-the-immersion-cube-and-the-immersive-learning-brain/269645196).

The assistant guides users through documenting immersive learning cases by: a) Prompting for detailed descriptions; b) Structuring information according to the provided frameworks; c) Cross-referencing case details with educational practices and strategies; d) Generating visual representations; and e) Producing a formatted document.

We first drafted a case description and then used this ILCS Assistant GPT to interactively analyze the case, checking its insights and analyses to determine which aspects were being omitted or misinterpreted. We then rewrote the case description to include those aspects or rephrased them and reapplied the ILCS Assistant GPT. This interactive improvement continued until we felt the outcome accurately reflected the case, which is presented below.

### IV. ANCIENT GREEK TECHNOLOGY – AN IMMERSIVE LEARNING CASE SHEET

This section presents the final ILCS that is the outcome of the method described earlier (Table 1). In a previous work, this scenario was employed in a comparison study between desktop VR and immersive VR [3].

TABLE I. THE IMMERSIVE LEARNING CASE SHEET FOR THE ANCIENT GREEK TECHNOLOGIES IN VR CASE.

| Case: Educating about Ancient Greek Technologies in Virtual Reality |
|---|
| **Description:** This case explores an immersive learning environment designed for Higher Education students to interact with two Ancient Greek technological artifacts: a) the Phyctories, a visual signal communication system based on torches arranged to represent letters of the Greek alphabet, and b) the Aeolosphere, a primitive steam engine consisting of a sphere with two curved nozzles attached to pipes connected to a boiler filled with water. When the water was heated, steam was produced, escaping through the nozzles and causing the sphere to rotate. Using Meta Quest 2 Head-Mounted Displays and their motion controllers with the VRChat platform, students interact to manipulate virtual objects (Active Learning Theories) and collaborate with peers (Foster collaboration and social activities), to learn about these historical technologies. The professor provides live guidance during the activities, helping students to adjust their actions in real time (Coaching, demonstrating and/or observing instruction). The VRChat platform supports learning through auditory, visual, and kinesthetic modes, engaging multiple senses in the learning process (Multimedia learning theories, Learning design for multimodal information). |
| The learning experience is structured around five core activities (Instructional design): |
| • **Introductory Presentation:** The professor delivers a presentation on the Phyctories system, followed by a discussion to address student questions (see Fig. a) (Reproduce traditional teaching practices in 3D). |
| • **Phyctories System Simulation:** Students interact with a virtual Phyctories system, using motion controllers to manipulate torches and represent letters of the Greek alphabet by consulting a table displaying the correspondence between Greek alphabet letters and torch positions (see Fig. b) (Interactive visualization, Information visualization and inference). |
| • **Group Collaboration:** Students are divided into groups to encode and decode messages using separate Phyctories systems (see Fig. c) (Foster collaboration and social activities). |
| • **Aeolosphere Simulation:** Teacher discusses the parts of the Aeolosphere engine (Expositional) and students light a virtual fire under a complete simulation of an Aeolosphere engine, observing its motion as steam is generated (see Fig. d) (Contextual theories). |
| **Final Reflection:** The session concludes with a discussion on both artifacts, allowing students to reflect on their experiences (Formative assessment). |

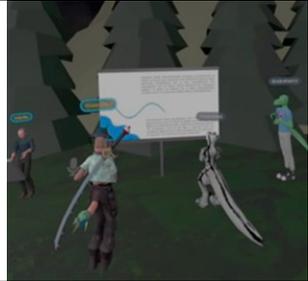
(a)

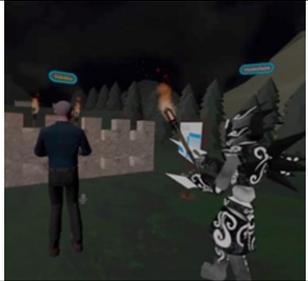
(b)

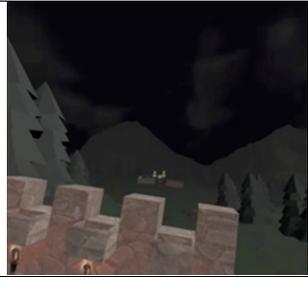
(c)

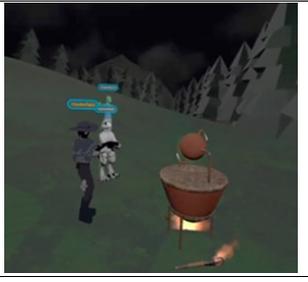
(d)

| Strategies tags | Practices tags |
|---|---|
| **Active Context cluster** | |
| Active Learning theories, Contextual theories | |
| **Engagement and Scaffolding cluster** | |
| Formative assessment, Instructional design | Coaching, demonstrating and/or observing instruction |
| **Real and Virtual Multimedia Learning cluster** | |

| Case: Educating about Ancient Greek Technologies in Virtual Reality | |
|---|---|
| Multimedia learning theories | Information visualization and inference, Learning design for multimodal information |
| **Presence cluster** | |
| Interactive visualization | |
| **Collaboration cluster** | |
| Collaborative learning | Foster collaboration and social activities |
| **Traditional Practices cluster** | |
| Expositional | Reproduce traditional teaching practices in 3D |
| **Proximal Uses** | **Immersion Cube Coordinates** System: 0.9 Narrative: 0.6 Agency: 0.65 |

## V. Discussion

Comparing the initial Ancient Greek Technology learning case description [3] with the one developed in this work, it is evident that employing the ILCS led to a more structured, comprehensive framework that enhanced the clarity and consistency of the case details. The co-intelligent use of the ILCS Assistant supported this process, since it prompted the research team to elaborate on pedagogical objectives, participant roles, and transitions between learning activities, enriching the descriptive quality beyond the initial description. By incorporating categories such as practices, strategies, proximal uses, and Immersion Cube Coordinates, this structured approach—guided by the ILCS assistant's iterative feedback—refined the case description to improve its coherence and comparability. This highlights the value of AI tools in producing rigorously standardized and analytically enriched immersive learning cases descriptions.

Under the interpretative lens of the ILCS, several avenues emerge for interpreting and discussing this case. Firstly, we have a structured perspective on the educational practices and strategies involved, leverageable when contrasting this case with others. Secondly, we notice several strategies without matching practices. These gaps are reflection opportunities: would it make sense for such isolated strategies to be supported or complemented by practices in those clusters? Or should the case be rethought not to rely on strategies that are not complemented by matching practices? The framework does not recommend "yes" nor "no" answers, merely exposes these opportunities for structured reflection upon the case.

On the Immersive Learning Cube part of the ILCS, this case is near to the conceptual location of the uses "Logistics" (overcoming resource shortcomings) and "Simulate the physical world". While the latter matches the current description, the former prompts another educational purpose reflection: the case could be planned, from a pedagogical integration perspective, to overcome shortcomings of access to these artifacts, to their frequency of use, or to its opportunity (for instance, since torches and boiling steam are not real, their interaction does not require safety supervision) or other occasions. There would be little effort in adapting the case to this use. On the other hand, adapting it to the farthest uses would require more significant planning changes.

Beyond these insights that the ILCS provides towards the case, the interaction with the Immersive Learning Case Sheet assistant revealed key impacts on the process of documenting and analyzing immersive learning cases. Initially, the Assistant highlighted the necessity for comprehensive and detailed case descriptions: the preliminary descriptions lacked specificity in areas such as participant roles in each learning experience, initiation processes, instructional guidance, and transitions between activities. The Assistant's feedback indicated that without this detailed information, the analysis might not fully capture the essence of the learning case.

This iterative process underscored the ILCS Assistant's role in refining documentation by acting as a co-worker that prompts for missing elements crucial for accurate analysis. Revising the case descriptions to include these specifics led us to more precise alignment with the ILB and ILC frameworks.

Engaging with the ILCS Assistant facilitated a step-by-step analysis using the frameworks. Upon requesting explicit references to strategies and practices, the ILCS Assistant incorporated these elements, enriching the analysis. Mapping the case onto the Immersive Cube allowed for categorization based on established educational uses, ensuring accurate classification when provided with standardized categories.

Moreover, the ILCS Assistant enabled us to attain a deeper understanding of the ILB & ILC concepts, by leading us to engage more critically within the bounds of the frameworks, rather than ad libitum discussion. Its interactive nature allowed for clarification of complex aspects through dialogue, contributing to our comprehension and analytical work.

Importantly, the ILCS Assistant not only aided in analyzing the current case but also highlighted areas for improvement in future case descriptions. Recognizing the limitations of initial descriptions informed a more meticulous approach in documenting learning experiences, which is essential for producing comparable and replicable studies.

However, the ILCS Assistant deviated from its knowledge base, both in format and classification aspects. At the format level, it blended the ILCS's foreseen space for lists of strategies and practices into a single mixed list area, and included a justification for each item, whereas the ILCS method prescribed that the original description should attest the presence of those items, not the list. At the content level, it modified the phrasing of the Strategies and Practices items, for instance by stating "Active learning" instead of "Active learning theories" (the ILB framework item name) or a single "Multimodal learning" item, one that does not exist in the ILB, rather mixing two items that should have been proposed: the strategy "Multimedia learning theories" and the practice "Learning design for multimodal information". We had to detect and correct those erroneous proposals.

## VI. Conclusion

The experiences with the ILCS Assistant in co-intelligent interaction suggest it holds significant potential to contribute to the field of immersive learning. By supporting standard approaches to the textual-visual documentation and analysis process, it addresses researchers' common challenges: variability in interaction fidelity and narrative elements among different cases. Given the ILCS Assistant frequent deviation from guiding frameworks' terminology and their reporting format, care must still be taken by research teams in verifying compliance and accuracy. This is to be expected from co-intelligent work paradigms [4], [5]: an LLM-based tool's usefulness is derived from the emergence of quality and inspiration from the dynamic interaction and critical appraisal by researchers, not as a shortcut or mere labor-saving tool. For instance, considering a scenario of analysis of multiple immersive training cases, for instance for addressing fire incidents or cultural heritage tours, the ILCS Assistant could streamline the comparison process by providing researchers documenting those cases with a consistent pull towards the framework categorization - even if its suggestions are tangent to the framework's items. If researchers exercise their critical thinking reliably, this will likely enhance the validity and reliability of findings by mitigating discrepancies arising from differences in equipment interaction and narrative contexts.

Contentiously, however, we hypothesize that even in situations where researchers do not entirely exert this verification of adherence to a framework, the interaction with the ILCS Assistant may contribute to ultimately improve reporting quality beyond the usual level. We should consider that multiple cases are often accompanied and reported by different researchers over time, with varying levels of understanding, background, and training (e.g., senior researchers from different fields, junior researchers, graduate students, research fellows and assistants). The usual reporting detail and quality is often of uneven quality. Inevitably, some human effort is lackadaisical, not entirely checking the adherence to the framework items. Therefore, by having a team employ the Immersive Learning Case Sheet paired with the ILCS Assistant, the constant pull can support more standardized and consistent descriptions. Thus, even in under-optimal conditions, this process may facilitate effective comparisons between related works and new learning cases, and even more so in optimal conditions of rigorous application. Systematically applying the ILB and ILC frameworks may therefore aid in identifying key similarities and differences, contributing to a deeper understanding of the educational impact of such applications.

In conclusion, the Immersive Learning Case Sheet and the ILCS Assistant serve as a valuable instrument for researchers and practitioners in immersive learning. By promoting interactively more rigorous documentation and analysis, they enhance the quality of case descriptions and foster a deeper understanding of immersive learning frameworks. Such instruments have the potential to significantly advance the field, enabling more effective comparisons and fostering the development of best practices in immersive learning research. Ultimately, we encourage researchers in this field and others to pair traditional methods and instruments with co-intelligent assistants, in support of stronger communication and tighter coupling and contrasting of qualitative results and reports.

### Acknowledgment

The authors acknowledge that ChatGPT was employed as a co-intelligent tool throughout redaction, to revise style and language, synthesize, and as interactive debating tool.

### References

[1] L. Alfieri, T. J. Nokes-Malach, e C. D. Schunn, «Learning Through Case Comparisons: A Meta-Analytic Review», *Educational Psychologist*, vol. 48, n.º 2, pp. 87–113, abr. 2013, doi: 10.1080/00461520.2013.775712.

[2] M. Glette e S. Wiig, «The Headaches of Case Study Research: A Discussion of Emerging Challenges and Possible Ways Out of the Pain», *TQR*, mai. 2022, doi: 10.46743/2160-3715/2022.5246.

[3] V. Kasapakis, E. Fokides, A. Kostas, A. Agelada, D. Gavalas, e G. Koutromanos, «Virtual Reality for Synchronous Learning in Higher Education», em *Extended Reality*, vol. 15030, L. T. De Paolis, P. Arpaia, e M. Sacco, Eds., em Lecture Notes in Computer Science, vol. 15030. , Cham: Springer Nature Switzerland, 2024, pp. 249–257. doi: 10.1007/978-3-031-71713-0_17.

[4] E. Mollick, *Co-intelligence: living and working with AI*. New York: Portfolio/Penguin, 2024.

[5] E. Schlemmer e L. Morgado, «Inven!RA: Um contributo para plataformas alinhadas com a Transformação Digital na Educação», *RE@D - Revista de Educação a Distância e Elearning*, p. e202403 Pages, mai. 2024, doi: 10.34627/REDVOL7ISS1E202403.

[6] B. Takács, «How and why affordable virtual reality shapes the future of education», *International Journal of Virtual Reality*, vol. 7, n.º 1, pp. 53–66, 2008.

[7] G. Freeman e D. Maloney, «Body, Avatar, and Me: The Presentation and Perception of Self in Social Virtual Reality», *Proc. ACM Hum.-Comput. Interact.*, vol. 4, n.º CSCW3, pp. 1–27, jan. 2021, doi: 10.1145/3432938.

[8] E.-E. Papadopoulou, D. Kavroudakis, A. Agelada, N. Zouros, N. Soulakellis, e V. Kasapakis, «Virtual reality in geoeducation: the case of the Lesvos Geopark», *Interactive Learning Environments*, pp. 1–17, jul. 2024, doi: 10.1080/10494820.2024.2374399.

[9] V. Chheang *et al.*, «Advanced liver surgery training in collaborative VR environments», *Computers & Graphics*, vol. 119, p. 103879, abr. 2024, doi: 10.1016/j.cag.2024.01.006.

[10] L. Morgado *et al.*, «Recommendation Tool for Use of Immersive Learning Environments», *2022 8th International Conference of the Immersive Learning Research Network …*. 2022.

[11] M. Castelhano, L. Morgado, e D. Pedrosa, «Instructional design models for immersive virtual reality: a systematic literature review», em *Atas do XXV Simpósio Internacional de Informática Educativa*, M. do R. Rodrigues, M. Figueiredo, e J. Torres, Eds., Setúbal, Portugal: Instituto Politécnico de Setúbal, 2023, pp. 272–278. Acedido: 16 de fevereiro de 2024. [Em linha]. Disponível em: https://repositorioaberto.uab.pt/handle/10400.2/15232

[12] D. Beck, L. Morgado, e P. O'Shea, «Educational Practices and Strategies with Immersive Learning Environments: Mapping of Reviews for using the Metaverse», *IEEE Trans. Learning Technol.*, vol. 17, pp. 319–341, 2024, doi: 10.1109/TLT.2023.3243946.

[13] D. Beck e L. Morgado, «Describing and Interpreting an Immersive Learning Case with the Immersion Cube and the Immersive Learning Brain», em *iLRN 2024 proceedings*, Glasgow, Scotland, in press.

[14] N. C. Nilsson, R. Nordahl, e S. Serafin, «Immersion Revisited: A review of existing definitions of immersion and their relation to different theories of presence», *Human Technology*, vol. 12, n.º 2, pp. 108–134, nov. 2016, doi: 10.17011/ht/urn.201611174652.

[15] D. Beck, L. Morgado, e P. O'Shea, «Finding the Gaps about Uses of Immersive Learning Environments: A Survey of Surveys», *Journal of Universal Computer Science*, vol. 26, n.º 8, pp. 1043–1073, 2020.